\pdfoutput=1

\documentclass[11pt]{article}

\usepackage[final]{acl}
\usepackage{times}
\usepackage{latexsym}
\usepackage[T1]{fontenc}
\usepackage[utf8]{inputenc}
\usepackage{microtype}
\usepackage{inconsolata}
\usepackage{graphicx}
\usepackage{multirow}
\usepackage{makecell}
\usepackage{booktabs}

\newcommand\blfootnote[1]{%
  \begingroup
  \renewcommand\thefootnote{}\footnote{#1}%
  \addtocounter{footnote}{-1}%
  \endgroup
}

\title{Structured Object Language Modeling (SoLM):\\Native Structured Objects Generation Conforming to Complex Schemas with Self-Supervised Denoising}

 \author{\\ {\bf Amir Tavanaei \quad Kee Kiat Koo \quad Hayreddin Ceker \quad Shaobai Jiang \quad  Qi Li \quad Julien Han} \AND {\bf Karim Bouyarmane} \\ \\
         Amazon, Seattle, USA \quad \url{https://so-lm.github.io}
}

\begin{document}
\maketitle

\begin{abstract}
In this paper, we study the problem of generating structured objects that conform to a complex schema, with intricate dependencies between the different components (facets) of the object. The facets of the object (attributes, fields, columns, properties) can be a mix of short, structured, type-constrained facts, or long natural-language descriptions. The object has to be self-consistent between the different facets in the redundant information it carries (relative consistency), while being grounded with respect to world knowledge (absolute consistency). We frame the problem as a Language Modeling problem (Structured Object Language Modeling) and train an LLM to perform the task natively, without requiring instructions or prompt-engineering. We propose a self-supervised denoising method to train the model from an existing dataset of such objects. The input query can be the existing object itself, in which case the model acts as a regenerator, completing, correcting, normalizing the input, or any unstructured blurb to be structured. We show that the self-supervised denoising training provides a strong baseline, and that additional supervised fine-tuning with small amount of human demonstrations leads to further improvement. Experimental results show that the proposed method matches or outperforms prompt-engineered general-purpose state-of-the-art LLMs (Claude 3, Mixtral-8x7B), while being order-of-magnitude more cost-efficient.
\end{abstract}

\section{Introduction}
\label{sec:intro}
Following\protect\blfootnote{Correspondence: \{atavanae,kiatkoo,hayro,shaobaij,qlimz, hameng,bouykari\}@amazon.com} natural-language text generation and code generation by the state-of-the-art Large Language Models (LLMs)~\cite{jiang2024mixtral,reid2024gemini,floridi2020gpt,claude2024,jiang2023mistral}, structured objects generation, also known as JSON (JavaScript Object Notation) generation or key-value pairs object generation, is a challenging problem for existing LLMs \cite{kitouni2024disk}.
It is one of the most desired behaviour of LLMs when used in production settings beyond traditional chatbot applications. It allows LLMs to be used as autonomous agents that integrate seamlessly with APIs, since JSON is the de-facto communication standard between APIs, and the universal string serialization format of structured objects. It also allows to use LLM outputs directly without any post-processing required, for example writing the output directly to a data store or passing it as input to subsequent functions. Finally, it allows to optimize LLM inference cost and number of calls and ensures self-consistency of the output by generating the entire object in a single LLM call, instead of generating each field of the object independently by an LLM query. 

General-purpose instruction-following and human-intent-aligned LLMs (chatbots) can be steered towards generating JSON objects outputs by specifying the requirement in their instruction prompts. Multiple prompting techniques have been tried to ensure that the output is a valid JSON that conforms to a schema, with variable success~\cite{beurer2024guiding,wang2024constraining,sengottuvelu2023jsonformer}. State-of-the-art LLM services such as OpenAI’s GPT-4, Anthropic's Claude 3, and Mistral AI's models, to name a few, have also recently introduced a “JSON-mode” that allows the user to steer the model’s output towards generating JSON, but without strict guarantee and still requiring the model to be explicitly instructed to output the JSON \cite{openai2024gpt4}.
Various wrapper libraries like JSONFormer~\cite{sengottuvelu2023jsonformer} allow to decompose the JSON generation problem into multiple independent value generation queries for each key, then using the generated values to fill the schema of the object in a post-processing re-composition step. The drawbacks of the approaches mentioned above are 1) long prompts/instructions and extensive prompt-engineering process, 2) unstable LLM's behaviour in response to prompt changes, 3) prerequisites for prompt preparation such as structured objects schema or ``keys'' in JSON, and 4) computationally expensive LLMs. 

To address these issues, we propose a self-supervised learning model to learn a native JSON Language Model, or \textbf{Structured object Language Model (SoLM)}, that natively generates objects that conform to a given structure (schema, class, database model, relational model, API specification, etc). The proposed SoLM acts as an object generator, but also as an object self-regeneration machine. No instructions or prompt-engineering is required for the model, which intelligently and autonomously understands what is the best possible schema and output given the input payload. Our model can also inherently perform multiple enhancement tasks while (re)generating the object. Tasks include 1) creation of the structured object from unstructured noisy input, 2) auto-completion of incomplete structured input, 3) error detection and auto-correction of noisy structured input, 4) auto-normalization of noisy structured input to desired normalization schemes, 5) auto-dependency resolution and auto-enforcement of inter-dependent parts/facets of the object.

In this work, we focus specifically on complex multi-facet objects with intricate dependencies between the different components (facets) of the object. The facets of the object (also known as attributes, fields, columns, properties) can be a mix of short, structured facts, or long, complex, natural-language descriptions. This type of structure naturally occurs in complex production use-case. Examples include product listings in online stores, house listings, job listings, entity records, etc. The object has to be self-consistent between the different facets and redundant information it carries (relative consistency), while being grounded and consistent with respect to a world knowledge about the entity (absolute consistency). We use an online store product catalog as an example application. For these types of e-commerce listings, some parts of the structure (e.g. title, product description, feature bullets, etc) are free-form natural language type of content, while other parts (structured metadata) are short form data-type and enumeration-constrained type of content. The proposed Structured Object Language Model handles the interleaving of these different types of content and ensures self-consistency between the natural-language portions and the structured content portions. 

Starting from a general purpose 7B parameter pre-trained Language Model, we first train our Structured Object Language Model using novel targeted denoising functions in a self-supervised manner (SoLM Self Supervised). The model is then further fine tuned based on few human generated high quality human demonstrations to align the LLM to human preferences (SoLM SFT: Supervised Fine Tuning). We compare this approach against prompt-engineering of SOTA LLMs, namely Claude 3.0 Sonnet and Mixtral-8x7B-Instruct, using two different prompt-engineering paradigms (whole object generation versus individual attributes generation). Results show that the proposed SoLM model is able to match the performance of prompt-engineered Claude 3.0 Sonnet while being order of magnitude more cost-effective.

\section{Self-Supervised Training}
\label{sec:ssft}
\begin{figure}[t]
    \centering
    \includegraphics[scale=0.25]{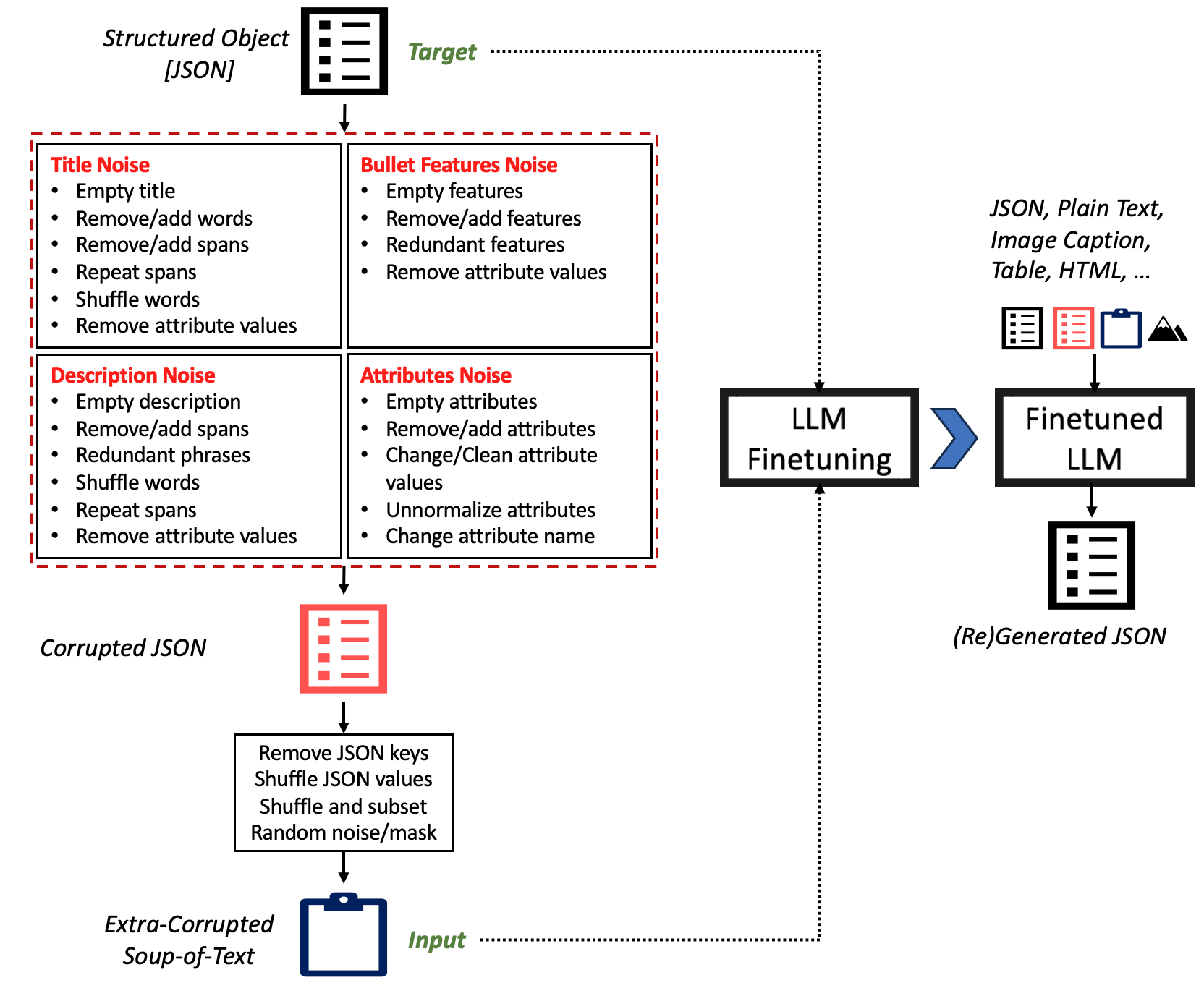}
    \caption{The noising functions applied to structured objects of products in e-commerce.}
    \label{fig:noising}
    \vspace*{-\baselineskip}
\end{figure}

In the following, we use the pre-trained MPT-7B as the backbone transformer architecture. MPT-7B is a decoder-only transformer pre-trained on English text and code including 1 trillion tokens \cite{MosaicML2023Introducing}. However, the proposed approach can be applied to any generative model (encoder-decoder or decoder-only). MPT-7B supports ALiBi position encoding for long text processing and generation regardless of the training text length and Flash Attention \cite{dao2022flashattention} for less GPU memory usage and a faster attention algorithm. 

The self-supervised learning approach does not require human labeled data and uses denoising techniques on a corpus of existing noisy data and tasks~\cite{tay2022ul2,raffel2020exploring}. The proposed noising functions are designed to minimize hallucination while preserving most of the pre-trained model's world knowledge. 

We define two modes of operations: one where the output has to strictly be grounded in data explicitly mentioned in the input payload/context, and one where the model is encouraged to rely on its inductive biases to guess the most plausible values even if not explicitly mentioned in the input payload. Those two modes of operation allow to cover a range of applications of structured objects generation or regeneration. Sec~\ref{sec:results} shows use-cases for the two modes.

\begin{figure}
    \centering
    \includegraphics[scale=0.23]{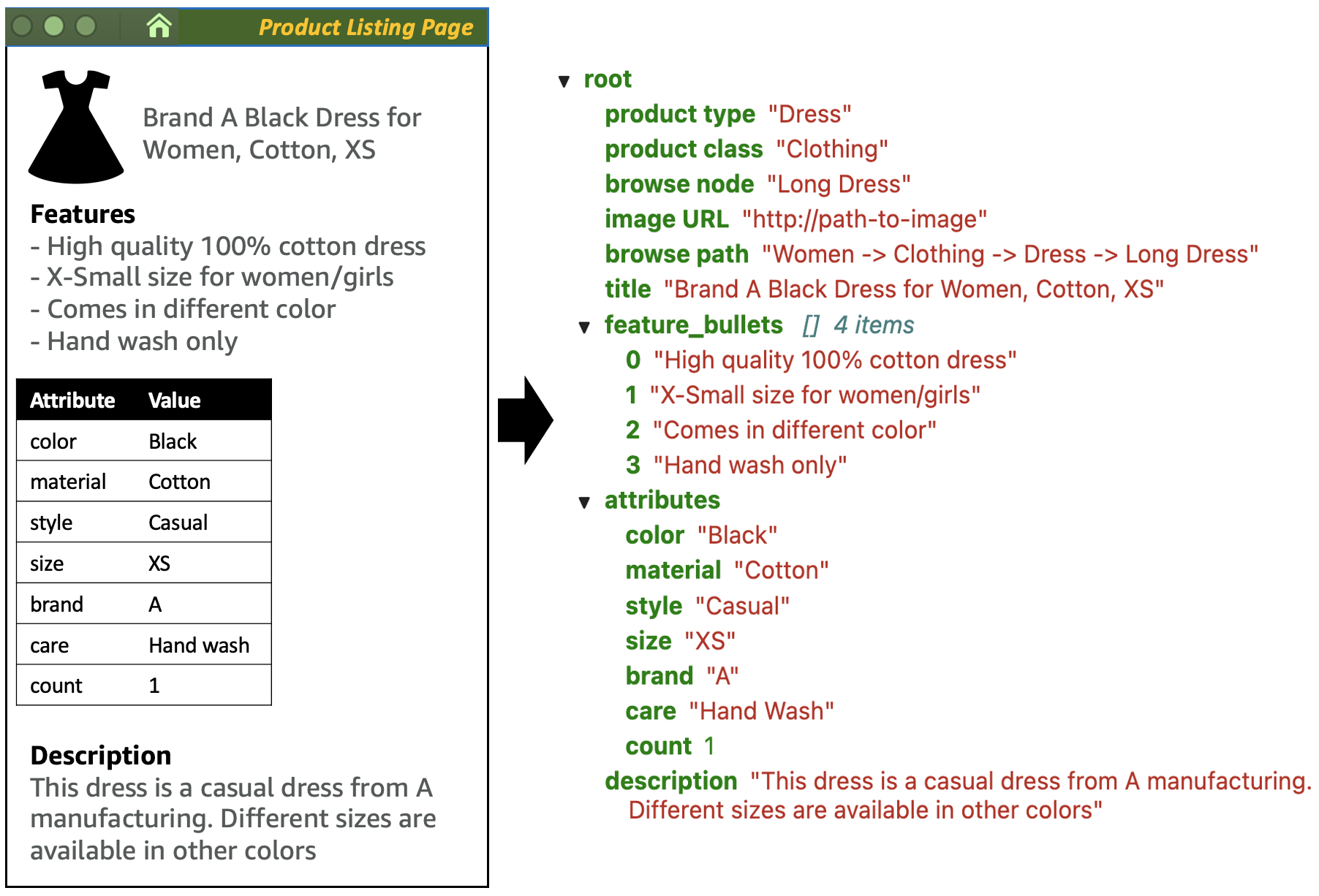}
    \caption{A structured object representing a product listing that includes multiple correlated components.}
    \label{fig:sample}
    \vspace*{-\baselineskip}
\end{figure}

\subsection{Noising Functions}
The two main components of the self-supervised learning stage are 1) a dataset of existing objects of interest (possibly noisy) and 2) a set of noising functions to apply on these noisy objects and generate very noisy inputs. The self-supervised model learns to remove the noise from the very noisy objects to recover their less noisy version. The assumption we make is that most of the objects in the dataset naturally carry some minimum amount of quality. We use these objects as target samples, and use the set of engineered noising functions to corrupt these target objects to form artificial input samples. The concatenation of the corrupted input and original target form one learning sample. 

Each component (facet) of an object is corrupted based on a subset of noising functions only targeting that component while using other components for noise customization. For instance, a structured object in an e-commerce catalog dataset would include four components: 1) title, 2) free-form bullet points describing the product features, 3) long description, and 4) tabular attributes such as color, material, brand, and size (up to a few hundred attributes per product, with schema depending on the product category). Each component becomes noisy by semi-randomly removing, changing, or adding information to the component's details. The ``semi-randomly'' here refers to random, controlled noises such that the noisy part of the component should be able to be recovered to the correct/complete format based on information understood from other components. For example, if the noising function removes or changes the color of a product in the tabular attributes, the correct color value should be mentioned (explicitly or implicitly based on the product category) in the other three components. This controlled noising function makes sure that the model does not hallucinate at inference and only generates or changes the texts if there are valid references in the whole structured object. This control can be tuned or turned off depending on the mode of operation (creative generation versus strictly grounded generation).

\subsection{Training Data Preparation}
For each structured object, a random combination of the noising functions explained above is calculated on-the-fly and applied during training.
The noising function for each component is randomly selected from a noising functions pool for that component. The noise intensity for each function (for instance, average number of words removed from the title) is itself randomly chosen from 0 to 100\%. At the end, a combination of noising functions on all the components are applied to the structured object to prepare a noisy structured object. 

While adequate for the use-case of regeneration of existing structured object for the purpose of cleaning (completing, correcting, normalizing, etc), the combination of the targeted noising functions mentioned above is not sufficient for the use-case of structured object generation from scratch given free-style input contexts or completely unstructured blurb inputs. To make the model more general and able to convert any informative text to the structured object of interest, with $\rho$ probability (e.g. $\rho=30\%$), we apply additional extreme noising to convert the corrupted JSON to so called ``soup-of-words'' (including complete structure destruction of the input and random shuffling of the tokens). Fig.~\ref{fig:noising} shows the noising functions and the final noise combination specified for e-commerce training data preparation. As shown in this figure, the model can get any input (plain text, structured object/JSON, image caption, tabular data, etc.) and generate a structured object/JSON with consistent, correct, and complete components in single pass.

\subsection{Denoising Training}
The fine-tuning in the decoder-only model is conducted by feeding the prepared input followed by the target structured object to the LLM for CLM (causal language modeling) training. The training sample template is

\begin{small}
\begin{verbatim}
   <BOS><input text><\n><target JSON><EOS>
\end{verbatim}
\end{small}

At inference, the model requires the noisy (original) structured object or text followed by ``\verb|\n|'' to generate the corresponding structured object.

\section{Supervised Fine Tuning}
\label{sec:sft}
Through the use of denoising functions, the self-supervised denoising trains the LLM to generate structured objects that conform to the target schema.
Although it is effective in adapting a pretrained LLM to the desired domain, as shown in Section \ref{sec:results}, the resulting model performance is limited as it is not explicitly trained to generate human preferred objects, especially on the subjective parts of the object (e.g. long description and free-form bullets).

Supervised Fine Tuning (SFT) is commonly used in the literature to align an LLM to desirable user responses \cite{ouyang2022training}.
Given a structured object, there exist a notion of human desirable or preferred responses.
The desired output is one that contains all relevant and factual information
representing the data.
The key to SFT lies in a demonstration dataset generated by human experts.

\section{Data Corpus}
\label{sec:data}
We validate our approach in the domain of e-commerce structured product catalog data.

\subsection{Self-Supervised Denoising Dataset}
We used a sample dataset from an established e-commerce online store containing 30 million product listings across thousands of product categories, filtered with simple heuristics to ensure a minimum data quality bar.
The components representing a product are correlated and have different data types like image, free-form text, structured attributes, and class names. These product listings are used as the target text generated by our model and the input is created by the proposed targeted noising functions explained in the previous section. The self-supervision stage maximizes for training data quantity over quality to ensure that the model is able to adapt to different products in the universe. 
Fig~\ref{fig:sample} shows an example of the structured object (JSON) representing a product in our dataset.

\subsection{SFT Dataset}
A naive approach to SFT would be to collect a small amount set of supervised training data for all product categories. However given the number of categories present in the data-set, this naive approach is both expensive and impractical to implement at scale. Instead, we propose a training pipeline similar to a funnel, where as it progress down the funnel the quality of the training data improves but with lower quantity:
\begin{itemize}
  \item \textbf{SFT Stage 1 - Existing High Quality Structured Objects}: We used an ad-hoc model trained to predict product quality to select existing structured objects from the noisy corpus which are of high quality for SFT training. This product quality model is trained based on existing business definition of product listing quality. The model is able to identify a sample subset of around 200K existing products which are deemed to be high quality according to this definition, filtered down from the original ~30M self-supervision dataset.
  \item \textbf{SFT Stage 2 - Human Labels}: Due to the aggressive nature of SFT Stage 1, many product categories remain under represented in the SFT training data set. Samples from these under represented products are then sent to human experts for labelling. Each human regenerated product is cross checked by another expert. This dataset contains around 3K product listings (structured objects).
  \end{itemize}

\section{Model Training and Evaluation Metrics}
\label{sec:modeltraining}
\subsection{Training}
The 7B SoLM model is trained on 5xAWS P4 instances, each with 8x 40GB A100 GPUs. We run ablations on a few backbone architectures including FLAN-T5 (XL and XXL) \cite{FlanT5}, MPT-7B \cite{MosaicML2023Introducing} and Mistral-7B~\cite{jiang2023mistral} to find the best pre-trained base model for the rest of the developments. See Appendix~\ref{sec:training}.

\subsection{Evaluation Metrics}\label{sub:eval}
The evaluation metrics we use in this paper fit the multi-facet structured objects (such as JSON data of e-commerce product listings). The evaluation metrics are divided into two categories: 1) metrics for free-form texts (title, feature bullets, and description) and 2) metrics for tabular attributes.

\textbf{Generated Free-form Facets} - The free-form texts are evaluated using subjective and objective methods. The objective evaluation is formulated by Rouge scores~\cite{lin2004rouge} on the reference texts for synthetically noised inputs. A more reliable evaluation is the subjective evaluation on the generated texts given original, real inputs. A text is labeled as ``correct'' if it 1) uses fluent language, 2) includes necessary attributes (e.g. color for shirt), 3) has no hallucination and false claims, and 4) represents the product based on the input information.

\textbf{Generated Tabular Attributes} - The aim of the proposed model is to generate correct, complete, and normalized objects, we define the two metrics:
\begin{itemize}
    \item \textbf{Correctness Rate / Precision}: Measures the number of correctly generated attributes divided by the number of generated attributes
    \item \textbf{Completeness Rate / Recall}: Measures the number of generated attributes divided by the number of required attributes
\end{itemize}

\section{Experiments and Results}
\label{sec:results}
\subsection{Offline Evaluations}
We performed initial experiments to assess the performance of the self-supervised model in comparison with a SOTA instruction-tuned LLM (Mixtral-8x7B-Instruct) with JSON-mode in zero-shot.
The self-supervised model outperforms zero-shot Mixtral significantly. For example on title generation, the self-supervised SoLM outperformed Mixtral by 40.38 percentage points on Rouge-L F1 Score.
Details are available in Appendix \textbf{A}.

As a proof of concept, Table~\ref{tab:composeres1} shows the results of the self-supervision model on the synthetic task of improving and regenerating the whole structured object in one pass after applying a combination of synthetic noises to all the object's components.

\begin{table}[h]
\scriptsize
\centering
\caption{Performance of our self-supervised Structured-Objects Language Model (Self Supervised SoLM) in denoising synthetically noised product listings.}
\begin{tabular}{lccc}
\toprule
\textbf{Object's Facet \emph{(eval. metric)}  }            & \textbf{Noised Inputs}  & \textbf{Regen. Outputs}\\ \midrule
Title \emph{(Rouge-L F1)}           &   52.09               &  \textbf{69.58}                 \\
Feature Bullets \emph{(Rouge-L F1)} &        64.67       &       \textbf{73.36 }              \\
Description \emph{(Rouge-L F1)}     &       54.45           &        \textbf{67.66 }        \\
Tabular Attributes \emph{(Accuracy)}  &  82.07   &   \textbf{90.32}\\ 
\bottomrule
\end{tabular}
\label{tab:composeres1}

\vspace*{-\baselineskip}
\end{table}

\subsection{Real Test Cases}
The real case benchmark consists of a sample of around 5K product listings randomly sampled from an e-commerce catalog, with the task of improving their quality and fixing any issue with the listing. The original (input) structured objects and the regenerated ones were all human labeled to measure the baseline versus the regenerated quality. 

Table~\ref{tab:dec_benchmark} shows the product listings quality, versus the quality of regenerated ones by the proposed model (natively) and by SOTA LLM extensively prompt-engineered for the task. Claude 3.0 Sonnet was prompt-engineered for generating the object in one LLM call (single prompt). We also compare against an alternative prompting strategy consisting in running multiple independent LLM calls for each component/attribute of the object, generating the structured object one piece at a time by executing the prompt for each attribute separately. This strategy requires >100 LLM runs per object followed by post processing for recomposing the object. Due to its high throughput requirement (~100X throughput), we could not use Claude 3.0 for this approach, therefore we used a SOTA self-hosted open-source LLM, namely Mixtral-8x7B-Instruct. Note that both prompt-engineering approaches require product categories and corresponding product schema to be given as input. This requires running an upstream product category classification model and connecting the prompt with a product-category to product-schema mapping table. The precision and recall in Table~\ref{tab:dec_benchmark} represent the correctness and completeness scores of the generated attributes (Sec.~\ref{sub:eval}). The title quality is a composite score of human scores of overall quality, and automatic quality check of title length and restricted characters/phrases in the title. The feature bullets quality is assessed by heuristics rules. 

As our SoLM model is trained in 2 stages (self-supervised training followed by SFT), we report results for both stages.
As reported in Table~\ref{tab:dec_benchmark}, the self-supervised model shows high precision (correctness) for structured attribute generation. General self supervised denoising increases the hallucination rate as the model is trying to fill out the missing parts as much as possible which drops precision (correctness). However, in our proposed targeted denoising, we define specific control variables in the noising functions (as explained earlier) to minimize the hallucination, resulting in high precision/correctness. The SFT model shows significant improvement in the free-form facet of title, which carries a strong element of human preference.
\begin{table}
\centering
\scriptsize
\caption{Product listing regeneration applied on 5K real case test. Best in bold, second-best underlined. TQ: title quality. FBQ: feature bullets quality.}
\setlength{\tabcolsep}{3.5pt}
\begin{tabular}{lccccc}
\toprule
\textbf{Model}      & \textbf{Relative Cost}    & \textbf{Precision}      & \textbf{Recall}                & \textbf{TQ} & \textbf{FBQ}  \\ \midrule

\makecell[l]{Input dataset \\ (baseline)} & -- & 85.31 & 46.69 & 42.24 & 55.39 \\
SoLM Self-Supervised                       & \textbf{1X (1 run)}               & \underline{83.30}                  & 60.20               & 57.93       & \underline{98.62}                \\
SoLM SFT   & \textbf{1X (1 run)}               & 82.30                  & \underline{65.70}               & \textbf{72.99}       & \underline{98.62}                \\
\makecell[l]{Claude 3.0 Sonnet\\(single prompt)}& 6.8X  (1 run)            & \textbf{83.90}                  & \textbf{67.40}               & \underline{66.47}       & \textbf{99.67}        \\
\makecell[l]{Mixtral-8x7B-Instruct\\ (1 run per attribute)} & 2X (M runs) & 81.80 & 58.4 & 58.37 & 76.62 \\
\bottomrule
\end{tabular}
\label{tab:dec_benchmark}
\end{table}

\begin{table}
\centering
\scriptsize
\caption{Comparison between the different approaches}
\setlength{\tabcolsep}{2.5pt}
\begin{tabular}{lccccccc}
\toprule
\textbf{Model}     & \textbf{Size}  & \textbf{Cost} & \begin{tabular}[c]{@{}l@{}}\textbf{Prompt}\\ \textbf{tokens}\end{tabular}  & \begin{tabular}[c]{@{}l@{}}\textbf{Need} \\ \textbf{schema}\end{tabular} & \begin{tabular}[c]{@{}l@{}} \textbf{Need prod.}\\ \textbf{category}\end{tabular} & \begin{tabular}[c]{@{}l@{}}\textbf{Nb. of} \\ \textbf{runs}\end{tabular}\\
\midrule
SoLM (7B)  & \textbf{7B}  & \textbf{1X} & \textbf{None} & \textbf{No}  & \textbf{No} & \textbf{1} \\
Claude 3.0 Sonnet & \textgreater{}100B & 7X         & +2k     & Yes                                                           & Yes  & \textbf{1}                                             \\
Mixtral-8x7B-Instruct & 8x7B & 2X         & +2k     & Yes                                                           & Yes  & >100                                         \\
\bottomrule
\end{tabular}
\label{tab:claudevsssft}

\vspace*{-\baselineskip}
\end{table}

We ran extensive iterations on Claude prompts, experimenting with multiple prompting approaches and following the model provider's best practices. We worked with expert Claude prompt-engineers to craft the prompts. The final prompt is a long prompt (>2000 tokens) explaining all the requirements, listing the product category, the corresponding schema for each category, and additional control instructions. The average performance of the SoLM SFT model is comparable to the best resulting Claude performance while requiring approximately 7 times less computations without any pre-processing requirement (product category classification, attribute list by product category,  etc.).

In another experiment, around 2K product listings with various arbitrary schemas--different from our dataset schema--were selected to be converted to our target schema. As shown in Table~\ref{tab:ple2k}, Claude performed best, while our model performed closely without being explicitly trained on this task. 

\begin{table}[h]
\centering
\scriptsize
\caption{Arbitrary schema to target schema conversion (new product listing generation). Free form texts quality is reported by \textbf{F}eatures \textbf{B}ullet \textbf{Q}uality, title \textbf{Rel}evance, title \textbf{Corr}ectness, and title \textbf{Cons}istency. }
\setlength{\tabcolsep}{2.5pt}
\begin{tabular}{lcccccc}
\toprule
\textbf{Model}                   & \textbf{Precision}        & \textbf{Recall}  & \textbf{FBQ} & \textbf{Rel.} & \textbf{Corr.} & \textbf{Cons.} \\ \midrule
SoLM (7B) & \underline{75.8} & 44.2 & \underline{98.2}    &\underline{96.8}    & \underline{75.0}  & \underline{97.0}\\
Claude 3.0 Sonnet               & \textbf{76.1} & \textbf{57.0}    & \textbf{98.8}   & \textbf{97.3}   & \textbf{76.8}  & \textbf{98.7} \\
Mixtral-8x7B-Instruct          & 62.6          & \underline{57.6} & 88.0   & 92.7  & 64.5 & 96.5\\ 
\bottomrule
\end{tabular}
\label{tab:ple2k}

\vspace*{-\baselineskip}
\end{table}

The last real case experiment involves plain unstructured blurb text as the only source of input information. The user provides a short text and/or image(s) (that can be converted to text by captioning) and the model generates the product listing in JSON format. We used a real dataset 70 samples evaluated by human auditors to assess the generated texts relevancy, correctness, and consistency like above. See Table~\ref{tab:plesparsetext} for results, that are consistent with the other experiments results.

\begin{table}[h]
\centering
\scriptsize
\caption{Unstructured blurb to structured object. Our model is not trained explicitly for this task}
\begin{tabular}{lccc}
\toprule
\textbf{Model}    &  \textbf{Relevance} & \textbf{Correctness} & \textbf{Consistency}\\
\midrule
SoLM (7B)                    & \textbf{99.5} & \underline{66.0} & 96.4    \\
Claude 3.0 Sonnet               & \textbf{99.5} & \textbf{71.3}    & \textbf{100}    \\
Mixtral-8x7B-Instruct & 98.4 & 56.5 & \underline{97.8} \\
\bottomrule
\end{tabular}

\label{tab:plesparsetext}

\vspace*{-\baselineskip}
\end{table}

\subsection{Online A/B Tests}
Structured product data are mostly self-reported by individual retailers when listing on e-commerce product websites. Studies have shown that these self-reported data can be sparse and contain noisy facts \cite{cheng-etal-2023-tab}.
In this paper we use the proposed LLM to improve the product titles that is provided by retailers. 
Specifically given all relevant information provided by retailers when listing a product, our goal is to enhance the initial retailer provided product title in a manner in which will improve our buyers experience in discovering products relevant to their intent.
Improving our buyer's shopping experience consequently will also meaningfully improve our retailers products exposure. 

We use the enhanced title - as output by our LLM - to run an online A/B test against the existing retailer provided version in an English Language Store over a period of 2 weeks. 
Evaluation results show that our customers (buyers) prefer the title generated by our LLM compared to the existing retailer provided version overall.
Particularly the revised titles improved revenue (p-value=0.059) and increased the total units purchased (p-value=0.034).

\section{Conclusion}
\label{sec:conclusion}
This paper proposes a new approach to generate structured objects in a single pass without needing any prompt nor objects' schema. The Structured Object Language Model (SoLM) is trained using a novel self-supervised training method incorporating a combination of targeted noising functions to help create or improve structured objects with complete, correct, and normalized components. The self-supervised model is further fine-tuned on human labeled data to improve the quality of the free-form text components (SoLM SFT).

In future work, we will extend the training stages by incorporating reinforcement learning from human feedback (RLHF) and Preference Optimization (DPO) to better capture human preference.

\bibliography{referencefull}

\appendix
\section{Initial Results: Proof of Concept}
\label{sec:poc}
\begin{figure*}[ht]
    \centering
    \includegraphics[scale=0.3]{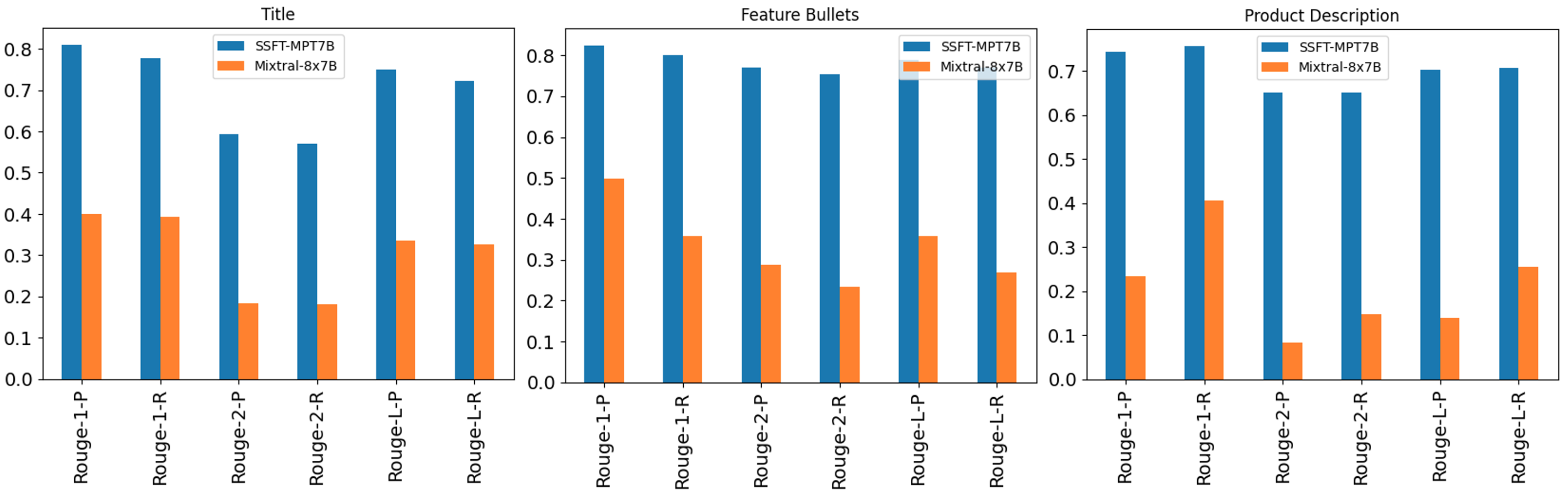}
    \caption{Zero-shot Mixtral with short prompt versus the self-supervised SoLM-7B for free-form text generation representing Title, Feature Bullets, and Product Description components of product listings in e-commerce. }
    \label{fig:basemixtral_vs_ssft}
\end{figure*}

For the initial experiments, we assess the performance of the self-supervised Structured Object Language Model (SoLM - 7B parameters) in comparison with the zero-shot Mixtral-8x7B-Instruct ($8\times 7$B parameters). The initial test dataset includes around 1K high quality product listings (structured objects) each including title, feature bullets, description, and a number of tabular attributes. To prepare the input, we applied diverse noises (that are also different from the noising functions used in training) on each component and expected the LLMs to improve and correct the noisy component. For free-form texts we used Rouge scores to compare the generated texts with the reference texts in the original structured objects. Since the self-supervised model is specifically trained on the structured objects, it's expected to perform better. Also, the zero-shot Mixtral does not have any information about the structured objects schema. Thus, only the Self Supervised SoLM is able to generate the list of tabular attributes that are missing in the structured object. 

Fig.~\ref{fig:basemixtral_vs_ssft} shows the Rouge scores of the zero-shot Mixtral versus SSFT for unstructured text generation. Table~\ref{tab:baseres11} compares the Self-Supervised SoLM and zero-shot Mixtral models in terms of 1) Rouge-L-F1 score, 2) subjective evaluation scores, and 3) tabular attributes accuracy. The subjective evaluation is conducted by assigning quality scores (0-100) to 150 generated texts by 2 human evaluators in which the highest scores is given to a text that is human readable and provides correct information about the structured object as explained in Section 5.2. The tabular attribute accuracy reports the correctness of the generated attribute values in comparison with the target attribute values in the original structured objects using fuzzy string matching.  
\begin{table}[ht]
\scriptsize
\centering
\caption{Initial comparison between the Self-Supervised Stuructured Object Language Model (SoLM) and zero-shot, short-prompted Mixtral-7x8B.}
\begin{tabular}{lcc}
\toprule
\textbf{Component Generation Eval}          & \textbf{SoLM-7B} & \textbf{Mixtral-8x7B} \\ \midrule
Title (Rouge-L F1)                 & \textbf{73.53}                                   & 33.15                                    \\
Feature Bullets (Rouge-L F1)       & \textbf{77.90}                                   & 30.68                                    \\
Description (Rouge-L F1)           & \textbf{70.50}                                   & 18.10                                   \\ \midrule 
Title (Subjective Score)           & \textbf{90.95}                                   & 70.68                                         \\
Feature Bullets (Subjective Score) & \textbf{91.76}                                    & 87.03                                           \\
Description (Subjective Score)     & \textbf{87.03}                                     & 85.27                                  \\ \midrule        
Tabular Attributes Accuracy        & \textbf{90.41}                                & NA     \\
\bottomrule
\end{tabular}
\label{tab:baseres11}
\end{table}

\section{Training Details}
\label{sec:training}
The FLAN-T5s are encoder-decoder models while others are decoder-only models. The current T5 architecture does not support existing Flash Attention algorithms~\cite{dao2022flashattention}. Thus, the training and inference time is expected to be high and the maximum input length for training must be less than 1024 tokens (due to memory and time complexity of the original self-attention on our EC2 instances). On the other hand, MPT and Mistral support Flash Attention which requires less memory and time and support very long input/output texts (>8k tokens). 

The models greater than 2B parameters cannot fit on a single GPU with $\leq$ 40 GB memory for training of long input/output texts (>1000 tokens). Thus, all the training codes are implemented by sharding the model using Pytorch-FSDP~\cite{zhao2023pytorch} to divide the computations cross multiple GPUs-Instances. The input/output maximum length were set to 1000 tokens for FLAN-T5-XL, 600 tokens for FLAN-T5-XXL and 8000 tokens for MPT and Mistral. Our catalog dataset including the components mentioned above has between 400 to 2500 tokens where make the T5-based models unable to learn long product listings. However, MPT and Mistral models can cover all the product listings (2500 input + 2500 output $\ll$ 8000) in our training dataset.

\end{document}